\documentclass[a4paper,10pt,twoside]{cpc-hepnp}
\usepackage{CJK,upgreek,fancyhdr}
\usepackage{multicol}
\usepackage{graphicx}
\usepackage{booktabs}
\usepackage{amssymb,bm,mathrsfs,bbm,amscd}
\usepackage[tbtags]{amsmath}
\usepackage{lastpage}
\usepackage{longtable}
\usepackage{changes}

\begin{document}
\begin{CJK*}{GB}{gbsn}

\fancyhead[c]{\small Chinese Physics C~~~Vol. xx, No. x (201x) xxxxxx}
\fancyfoot[C]{\small 010201-\thepage}


\title{$\mathcal{\alpha}$ decay properties of nucleus $^{297}$Og within two-potential approach
\thanks{Supported by the National Natural Science Foundation of China (Grants No. 11205083 and No. 11505100), the Construct Program of the Key Discipline in Hunan Province, the Research Foundation of Education Bureau of Hunan Province, China (Grant No. 15A159), the Natural Science Foundation of Hunan Province, China (Grants No. 2015JJ3103 and No. 2015JJ2121), the Innovation Group of Nuclear and Particle Physics in USC, the Shandong Province Natural Science Foundation, China (Grant No. ZR2015AQ007), Hunan Provincial Innovation Foundation For Postgraduate (Grant No. CX2017B536).}}

\author{%
     Jun-Gang Deng (邓军刚)$^{1}$
\quad Jun-Hao Cheng (程俊皓)$^{1}$
\quad Bo Zheng (郑波)$^{1;1)}$\email{zhengb@ihep.ac.cn}
\quad Xiao-Hua Li (李小华)$^{1,2,3;2)}$\email{lixiaohuaphysics@126.com}%
}
\maketitle

\address{%
$^1$ School of Nuclear Science and Technology, University of South China, Hengyang 421001, China\\
$^2$ Cooperative Innovation Center for Nuclear Fuel Cycle Technology $\&$ Equipment, University of South China, Hengyang 421001, China\\
$^3$ Key Laboratory of Low Dimensional Quantum Structures and Quantum Control, Hunan Normal University, Changsha 410081, China\\
}

\begin{abstract}
The $\mathcal{\alpha}$ decay half-life of unknown nucleus $^{297}$Og is predicted within two-potential approach, and $\mathcal{\alpha}$ preformation probabilities of 64 odd-$A$ nuclei in the region of proton numbers $82<Z<126$ and neutron numbers $152<N<184$ from $^{251}$Cf to $^{295}$Og are extracted. In addition, based on the latest experimental data, a new set of parameters for $\mathcal{\alpha}$ preformation probabilities considering the shell effect and proton-neutron interaction are obtained. The predicted $\mathcal{\alpha}$ decay half-life of $^{297}$Og is 0.16 ms within a factor of 4.97 as well as the spin and parity of ground states for nuclei $^{269}$Sg, $^{285}$Fl and $^{293}$Lv are $3/2^+$, $3/2^+$ and $5/2^+$, respectively.
\end{abstract}



\footnotetext[0]{\hspace*{-3mm}\raisebox{0.3ex}{$\scriptstyle\copyright$}2016
Chinese Physical Society and the Institute of High Energy Physics
of the Chinese Academy of Sciences and the Institute
of Modern Physics of the Chinese Academy of Sciences and IOP Publishing Ltd}%

\begin{multicols}{2}

\section{Introduction}

During several decades, the synthesis of superheavy nuclei has been a hot area of research in nuclear physics. Experimentally, elements 107--112 were synthesized in cold-fusion reactions at the separator for heavy-ion products (SHIP) facility at the GSI Helmholtz Centre for Heavy Ion Research in Germany \cite{RevModPhys.72.733,doi:10.1524/ract.2011.1854,doi:10.1063/PT.3.2880}. Through hot-fusion reactions between $^{48}$Ca beams and radioactive actinide targets, elements 113--118 were synthesized \cite{ PhysRevC.76.011601,PhysRevC.74.044602,0954-3899-34-4-R01,PhysRevLett.104.142502,PhysRevLett.105.182701,Hofmann2012}. In the future, the synthesis of $^{297}$Og is excepted to be via the reaction $^{249}$Cf+$^{48}$Ca$\to^{297}$Og at the Flerov Laboratory of Nuclear Reactions (FLNR) in Dubna, Russia \cite{ doi:10.1063/PT.3.2880}.
If experiment would be succeed, $^{297}$Og will be the nucleus observed with the largest number of neutrons 179 and  closest to predicted neutron numbers $N$=184 shell closure \cite{SOBICZEWSKI1966500,Mosel1969}.

Spontaneous fission and $\mathcal{\alpha}$ decay are the two main decay modes of superheavy nuclei. For superheavy nuclei around Rf, spontaneous fission is a stronger candidate compared to $\mathcal{\alpha}$ decay \cite{PhysRevC.94.054621}. For the majority of recently synthesized proton-rich superheavy nuclei, $\mathcal{\alpha}$ decay is the dominant decay mode because shell closure makes the superheavy nuclei stable against spontaneous fission \cite{PhysRevC.94.054621}. Recently, Bao $et\ al.$ also predicted the decay mode of $^{297}$Og is $\mathcal{\alpha}$ decay \cite{PhysRevC.95.034323}. $\mathcal{\alpha}$ decay, as an important tool to study superheavy nuclei, provides abundant information of nuclear structures and stabilities for superheavy nuclei. There are many theoretical models used to study the $\mathcal{\alpha}$ decay, such as the fission-like model, shell model, cluster model, and so on \cite{PhysRevLett.69.37,PhysRevC.51.559,0954-3899-26-8-305,PhysRevC.73.031301,PhysRevC.74.014304,PhysRevC.74.014312,PhysRevLett.103.072501,PhysRevC.77.054318,PhysRevC.81.064318,BAO201485}. The two-potential approach (TPA) \cite{PhysRevLett.59.262, PhysRevA.69.042705} was put forward to investigate the quasi-stationary problems, initially. Recently, it is widely used to deal with $\mathcal{\alpha}$ decay \cite{PhysRevC.85.027306, QIAN201182, QIAN20111,PhysRevC.93.034316,PhysRevC.94.024338,PhysRevC.95.044303, PhysRevC.95.014319,1674-1137-41-1-014102}. In our previous works \cite{PhysRevC.93.034316,PhysRevC.94.024338,PhysRevC.95.044303, PhysRevC.95.014319,1674-1137-41-1-014102}, we adopted TPA to systematically study $\mathcal{\alpha}$ decay half-lives of even-even, odd-$A$ and doubly-odd nuclei, and the calculations could well reproduce the experimental data.

The aim of this work is to predict the $\mathcal{\alpha}$ decay half-life $T_{1/2}$ of $^{297}$Og. Due to $T_{1/2}$ is sensitive to $\mathcal{\alpha}$ decay energy $Q_{\alpha}$, how to select a precise $Q_{\alpha}$ is one of the heart of the matters in predicting $T_{1/2}$ of $^{297}$Og. Theoretic predictions of $Q_\alpha$ for superheavy nuclei are performed by following mass models: M$\ddot{\text{o}}$ller $et\ al.$ (FRDM) \cite{MOLLER1995185}, Duflo and Zuker (DZ) \cite{PhysRevC.52.R23}, Nayak and Satpathy (INM) \cite{Nayak2012616}, Wang and Liu (WS3+) \cite{PhysRevC.84.051303}, Wang $et\ al.$ (WS4+) \cite{PhysRevC.93.014302,Wang2014215}, Muntian $et\ al.$ (HN) \cite{2001AcPPB..32..691M, Sobiczewski2007292}, Kuzmina $et\ al.$ (TCSM) \cite{PhysRevC.85.014319}, Goriely $et\ al.$ (HFB31) \cite{PhysRevC.93.034337}, and Liran $et\ al.$ (SE) \cite{PhysRevC.62.047301}. It is found that the WS3+ model \cite{PhysRevC.84.051303} is the most accurate one to reproduce the experimental $Q_\alpha$ of superheavy nuclei \cite{0954-3899-43-9-095106,PhysRevC.94.051302}. In order to accurately predict the half-life of nucleus of $^{297}$Og, we systematically investigate $\mathcal{\alpha}$ preformation probabilities of 64 odd-$A$ nuclei which $82<Z<126$ and $152<N<184$ from $^{251}$Cf to $^{295}$Og within TPA, the $\mathcal{\alpha}$ decay energy and half-lives are taken from the latest evaluated nuclear properties table NUBASE2016 \cite{1674-1137-41-3-030001} and evaluated atomic mass table AME2016 \cite{1674-1137-41-3-030002,1674-1137-41-3-030003} except the $Q_{\alpha}$ of nucleus $^{297}$Og is taken from WS3+ \cite{PhysRevC.84.051303}.

This article is organized as follows. In Section 2, the theoretical framework of calculating $\mathcal{\alpha}$ decay half-life is briefly described. The detailed calculations and discussions are presented in Section 3. Finally, a summary is given in Section 4.
\section{Theoretical framework}

In the TPA, the total interaction potential $V(r)$, between $\mathcal{\alpha}$ particle and daughter nucleus, is composed of the nuclear potential $V_N(r)$, Coulomb potential
$V_C(r)$ and centrifugal potential $V_l(r)$. It can be expressed as
\begin{equation}
\
V(r)=V_N(r)+V_C(r)+V_l(r)
.\label{subeq:1}
\end {equation}
The nuclear potential is determined within two-body model and assumed that $\mathcal{\alpha}$ particle preformed at surface of parent nucleus and the strong attractive nuclear interaction can be approximately replaced by a square well potential \cite{PhysRevC.87.024308, PhysRevC.95.014319}. In present work we choose a type of cosh parametrized form for the nuclear potential, which is obtained by analyzing experimental data of $\mathcal{\alpha}$ decay \cite{PhysRevLett.65.2975} and expressed as
\begin{equation}
\
V_N(r)=-V_0\frac{1+\mathrm{cosh}(R/a_0)}{\mathrm{cosh}(r/a_0)+\mathrm{cosh}(R/a_0)},
\label{subeq:1}
\end {equation}
where $V_0$ and $a_0$ are the depth and diffuseness of the nuclear potential. In our previous work \cite{PhysRevC.93.034316}, we have obtained a set of isospin dependent parameters, which is $a_0$=0.5958 fm and $V_0=192.42+31.059\frac{N_d-Z_d}{A_d}$ MeV, where $N_d$, $Z_d$ and $A_d$ are the neutrons, protons and mass number of daughter nucleus, respectively. $R$, the nuclear potential sharp radius, is empirically calculated within the nuclear droplet model and proximity energy \cite{0954-3899-26-8-305} with the mass number of parent nucleus $A$ and expressed as
\begin{equation}
\
R=1.28A^{1/3}-0.76+0.8A^{-1/3}
.\label{subeq:1}
\end {equation}
The Coulomb potential $V_C(r)$, obtaining under assumption of uniformly charged sphere with $R$, is expressed as
\begin{equation}
\
V_C(r)=\left\{\begin{array}{ll}

\frac{Z_dZ_{\alpha}e^2}{2R}[3-(\frac{r}{R})^2],&\text{{r}\textless{R}},\\

\frac{Z_dZ_{\alpha}e^2}{r},&\text{{r}\textgreater{R}},

\end{array}\right.
\label{subeq:1}
\end {equation}
where $ Z_{\alpha}=2$ represents the proton number of $\mathcal{\alpha}$ particle. In present work, we employ the Langer modified centrifugal barrier for $V_l(r)$, on account of $l(l+1){\to}(l+1/2)^2$ is a necessary correction for one-dimensional problems \cite{1995JMP....36.5431M}. It can be calculated by
\begin{equation}
\
V_l(r)=\frac{{\hbar}^2(l+1/2)^2}{2{\mu}r^2}
,\label{subeq:1}
\end {equation}
where $\mu=\frac{{m_d}{m_{\alpha}}}{{m_d}+{m_{\alpha}}}$ denotes the reduced mass between preformed $\mathcal{\alpha}$ particle and daughter nucleus with ${m_d}$ and $m_{\alpha}$ being the mass of daughter nucleus and $\mathcal{\alpha}$ particle. $l$ is the orbital angular momentum taken away by the $\mathcal{\alpha}$ particle. $l=0$ for the favored $\mathcal{\alpha}$ decays, while $l{\ne}0$ for the unfavored decays. Based on the conservation law of angular momentum \cite{PhysRevC.79.054614}, the minimum angular momentum $l_{\text{min}}$ taken away by the $\mathcal{\alpha}$ particle can be obtained by
\begin{equation}
\
l_{\text{min}}=\left\{\begin{array}{llll}

{\Delta}_j,&\text{for even${\Delta}_j$ and ${\pi}_p$= ${\pi}_d$},\\

{\Delta}_j+1,&\text{for even${\Delta}_j$ and ${\pi}_p$$\ne$${\pi}_d$},\\

{\Delta}_j,&\text{for odd${\Delta}_j$ and ${\pi}_p$$\ne$${\pi}_d$},\\

{\Delta}_j+1,&\text{for odd${\Delta}_j$ and ${\pi}_p$= ${\pi}_d$},

\end{array}\right.
\label{6}
\end {equation}
where ${\Delta}_j= |j_p-j_d|$, $j_p$, ${\pi}_p$, $j_d$, ${\pi}_d$ represent spin and parity values of the parent and daughter nuclei, respectively.

$\mathcal{\alpha}$ decay half-lives $T_{1/2}$, an important indicator for nuclear stability, is calculated by decay width $\Gamma$ or decay constant $\lambda$ and expressed as
\begin{equation}
\
T_{1/2}=\frac{{\hbar}ln2}{\Gamma}=\frac{ln2}{\lambda}
.\label{subeq:1}
\end{equation}
In framework of TPA, $\Gamma$ can be given by
\begin{equation}
\
\Gamma=\frac{{\hbar}^2P_{\alpha}FP}{4\mu}
,\label{subeq:1}
\end{equation}
where $P$ is penetration probability, namely, Gamow factor, obtained by the Wentzel-Kramers-Brillouin (WKB) method and expressed as
\begin{equation}
\
P=\exp(-2{\int_{r_2}^{r_3} k(r) dr})
,\label{subeq:1}
\end{equation}
where $ k(r)=\sqrt{\frac{2\mu}{{\hbar}^2}|Q_{\alpha}-V(r)|}$ denotes the wave number of the $\mathcal{\alpha}$ particle. $r$ is the center of mass distance between the preformed $\mathcal{\alpha}$ particle and daughter nucleus. $r_2$, $r_3$ and following $r_1$ are the classical turning points. They satisfy conditions $V (r_1) = V (r_2) = V (r_3) =Q_\alpha$.
The normalized factor $F$, denoting the assault frequency of $\mathcal{\alpha}$ particle, can be obtained by
\begin{equation}
\
F{\int_{r_1}^{r_2} \frac{1}{2k(r)} dr}=1.
\label{subeq:1}
\end{equation}
\label{11}
On account of the complicated structure of the quantum many-body systems, there are a few works \cite{QI201677,PhysRevC.77.054318,PhysRevC.92.044302,PhysRevC.93.011306,PhysRevC.95.061306} studying $\mathcal{\alpha}$ preformation probabilities $P_{\alpha}$ from the viewpoint of microscopic theory. Phenomenologically, the $\mathcal{\alpha}$ preformation probability $P_{\alpha}$ is extracted by
\begin{equation}
\
 P_{\alpha}=P_0\frac{T_{1/2}^{\text{cal}}}{T_{1/2}^{\text{exp}}},
\end{equation}
where $T_{1/2}^{\text{exp}}$ denotes experimental half-life. $T_{1/2}^{\text{cal}}$ represents the calculated $\mathcal{\alpha}$ decay half-life based on an assumption $P_{\alpha}=P_0$. In accordance with the calculations by adopting the density-dependent cluster model (DDCM) \cite{XU2005303}, $P_0$ is 0.43 for even-even nuclei, 0.35 for odd-$A$ nuclei, and 0.18 for doubly-odd nuclei, respectively. In present work $P_0$ is 0.35. Recently, the variation tendency of $P_{\alpha}$ can be estimated by the analytic formula \cite{PhysRevC.80.057301, PhysRevC.93.034316, GUO2015110,PhysRevC.95.044303, PhysRevC.95.014319}, which is considered the nuclear shell structure and proton-neutron interaction, and expressed as
\begin{equation}
\label{12}
\begin{aligned}
\log_{10}P_{\alpha}=&a+b(Z-Z_1)(Z_2-Z)+c(N-N_1)\\
&(N_2-N)+dA+e(Z-Z_1)(N-N_1),
\end{aligned}
\end{equation}
where $Z$ $(N)$ denotes the proton (neutron) numbers of parent nucleus. $Z_1$ ($N_1$), $Z_2$ ($N_2$) denote the proton (neutron) magic numbers with $Z_1<Z<Z_2$ and $N_1<N<N_2$.

\section{Results and discussions}

In our previous works\cite{PhysRevC.93.034316,PhysRevC.94.024338,PhysRevC.95.044303, PhysRevC.95.014319,1674-1137-41-1-014102}, we found that the behaviors of $\mathcal{\alpha}$ preformation probabilities of the same kinds nuclei(even-even nuclei, odd-$A$ nuclei and doubly-odd nuclei) in the same region, which is divided by the magic numbers of proton and neutron, can be described by Eq. (\ref{12}). For the purpose of a precise prediction for $^{297}$Og, we systematically study all of 64 odd-$A$ nuclei included odd $Z$, even $N$ (odd-even) and even $Z$, odd $N$ (even-odd) nuclei in the same region with $^{297}$Og from $^{251}$Cf to $^{295}$Og. For the odd-$A$ nuclei, excitation of single nucleon causes the high-spin isomers. Our previous works \cite{1674-1137-41-1-014102,PhysRevC.94.024338} indicate that both ground and isomeric states can be treated in a unified way for $\mathcal{\alpha}$ decay parent and daughter nuclei.

$l_{\text{min}}$ is an important input for calculating $T_{1/2}$, which can be calculated by Eq. (6). However we do not know the spin and parity values of parent nuclei and/or daughter nuclei for $\mathcal{\alpha}$ decay $^{271}$Sg, $^{271}$Bh, $^{273}$Hs and so on. In present work, the minimum angular momentums are approximatively taken as $l_{\text{min}}=0$ for those $\mathcal{\alpha}$ decay. In order to verify whether this assumption is right or not, we plot the logarithm deviation between $T^\text{pre}_{1/2}$ and $T^\text{exp}_{1/2}$ for those nuclei in Fig.~\ref{fig1}, where $T^\text{pre}_{1/2}=\frac{P_0T_{1/2}^{\text{cal}}}{P_{\alpha}^*}$  with $P_{\alpha}^*$ obtained by Eq. (\ref{12}) and parameters in Table \ref{table1} as well as $T^\text{cal}_{1/2}$ taking $P_{\alpha}=P_0$. From Fig.~\ref{fig1}, we can clearly see that the values of $\log_{10}T^\text{pre}_{1/2}- \log_{10}T^\text{exp}_{1/2}$ are , on the whole, around 0. It indicates that the spin and parity of ground states for those nuclei and their daughter nuclei may be equal. Therefore, the spin and parity of ground states for nuclei $^{269}$Sg, $^{285}$Fl, $^{293}$Lv may be $3/2^+$, $3/2^+$ and $5/2^+$, respectively. In addition, for $^{271}$Bh, $^{275}$Hs, $^{281}$Ds$^\text{m}$ and $^{295}$Og, the deviations between predictions and experimental data are slightly big. Those indicate that the spin and parity of ground states for the above four nuclei and their daughter nuclei are potentially different.

In the following, we calculate $\mathcal{\alpha}$ decay half-lives with $P_{\alpha}=P_0$ and extract corresponding $P_{\alpha}$ within Eq. (11). And then, we fit all of $P_{\alpha}$ based on Eq. (12) and extract relevant parameters given in Table \ref{table1}, where in  region of $82<Z\leq{126}$ and $152<N\leq{184}$, $Z_1=82$, $Z_2=126$, $N_1=152$, $N_2=184$. In our previous work \cite{PhysRevC.95.014319}, we have obtained a set of parameters for this region. The standard deviation $\sigma_\text{pre}=\sqrt{\sum ({\log_{10}T^\text{pre}_{1/2}-\log_{10}T^\text{exp}_{1/2}})^2/n}$ denotes deviations of $\mathcal{\alpha}$ decay half-live between predictions considering $\mathcal{\alpha}$ preformation probabilities correction and experimental data for those 64 odd-$A$ nuclei. The values of $\sigma_\text{pre}$ drops from 0.739 by adopting the parameters of past work \cite{PhysRevC.95.014319} to 0.696 by using the new ones, which indicates that predictions adopting new parameters are improved by $\frac{0.739-0.696}{0.739}=5.82\%$. The standard deviations $\sigma_\text{cal}=\sqrt{\sum ({\log_{10}T^\text{cal}_{1/2}-\log_{10}T^\text{exp}_{1/2}})^2/n}$ between calculated $T_{1/2}$ with $P_{\alpha}=P_0$ and experimental ones for those 64 odd-$A$ nuclei is 1.177, hence, by adopting $P_{\alpha}^*$ considering the shell effect and proton-neutron interaction, the standard deviation reduces $\frac{1.177-0.696}{1.177}=40.87\%$.

The detailed calculations are given in Table \ref{table2}. In this table, the first four columns are $\mathcal{\alpha}$ transition, $\mathcal{\alpha}$ decay energy, spin-parity transformation and the minimum orbital angular momentum $l_{\text{min}}$ taken away by $\mathcal{\alpha}$ particle, respectively. The fifth, sixth and seventh ones are the experimental half-life $T^\text{exp}_{1/2}$, calculated half-life $T^\text{cal}_{1/2}$ by TPA with $P_{\alpha} =P_0$ and extracted $\mathcal{\alpha}$ preformation probability $P_{\alpha}$ with Eq. (11), respectively. The last two ones are $\mathcal{\alpha}$ preformation probability $P_{\alpha}^*$ and predicted $\mathcal{\alpha}$ decay half-lives $T^\text{pre}_{1/2}$. From the Table \ref{table2}, we can find that for some nuclei, such as $^{255}$Md, $^{271}$Sg, $^{271}$Bh, $^{273}$Ds$^{\text{m}}$, $^{279}$Ds, $^{281}$Ds and so on, the extracted $P_{\alpha}$ are especially small. For the case of nuclei $^{271}$Sg, $^{271}$Bh and $^{279}$Ds, the reason is that our assumptions of their $l_{\text{min}}=0$ may be inappropriate. While, for the case of nuclei $^{255}$Md, $^{273}\mathrm{Ds}^{\text{m}}$ and $^{281}$Ds, it is that the uncertain and/or estimated spin and parity may be inaccuracy. The inaccuracy $l_{\text{min}}$ might cause the great differences between $T_{1/2}^{\text{exp}}$ and $T_{1/2}^{\text{cal}}$ as well as smaller $P_{\alpha}$.
The experimental data and predicted results are plotted as logarithmic forms in Fig.~\ref{fig2}. In this figure, the blue triangle and red circle represent the experimental half-lives $T_{1/2}^\text{exp}$, and predictions $T_{1/2}^\text{pre}$, respectively. From this figure, we can see that the predicted half-lives can well reproduce the experimental ones basically. For more intuitively, we plot the logarithms differences between predictions and experimental data in Fig.~\ref{fig3}. From this figure, we can clearly see that the values of $\log_{10}{T^\text{pre}_{1/2}}-\log_{10}{T^\text{exp}_{1/2}}$ are mainly around 0, indicating our predictions being in good coincidence with the experimental data. Therefore, extending our study to predict the $\mathcal{\alpha}$ decay half-life and $\mathcal{\alpha}$ preformation probability of nucleus $^{297}$Og may be believable. Then we calculate $T_{1/2}$ of $^{297}\text{Og}$ by $T^\text{pre}_{1/2}=\frac{P_0T_{1/2}^{\text{cal}}}{P_{\alpha}^*}$ with $P_{\alpha}^*$ in Table \ref{table2}, while $Q_{\alpha}$ is taken from WS3+ \cite{PhysRevC.84.051303} and $l_{\text{min}}=0$. Finally, according to the standard deviations $\sigma_{pre}$ for 64 odd-$A$ nuclei, in the same region with $^{297}\text{Og}$, is 0.696, then the predicted $\mathcal{\alpha}$ decay half-life of $^{297}$Og is 0.16 ms within a factor of 4.97.

\begin{center}
\tabcaption{ \label{table1} The parameters of $P_{\alpha}^*$ for odd-$A$ nuclei from $82<Z\leq126$ and $152<N\leq184$.}
\footnotesize
\begin{tabular*}{80mm}{c@{\extracolsep{\fill}}ccccc}
\toprule a&b&c&d&e\\
\hline
15.4694&-0.0054&-0.0014&-0.0546&0.0019\\
\bottomrule
\end{tabular*}
\vspace{0mm}
\end{center}
\vspace{0mm}

\end{multicols}

\begin{center}
\tabcaption{Calculations of $\mathcal{\alpha}$ decay half-lives and $\mathcal{\alpha}$ preformation probabilities and predicted half-lives. Elements with upper suffixes `m', `n' and `p' indicate assignments to excited isomeric states (defined as higher states with half-lives greater than 100 ns). Suffixes `p' also indicate non-isomeric levels, but used in the AME2016 \cite{1674-1137-41-3-030002,1674-1137-41-3-030003}. `()' means uncertain spin and/or parity. `\#' means values estimated from trends in neighboring nuclides with the same $Z$ and $N$ parities.}
\footnotesize
\label{table2}
\begin{longtable}{ccccccccc}
\hline {$\mathcal{\alpha}$ transition}&$Q_{\alpha}$ (MeV)&${j^{\pi}_{p}}\to{j^{\pi}_{d}}$ &$l_{\text{min}}$&$T^\text{exp}_{1/2}$ (s)&${T_{1/2}^\text{cal}}$ (s)&${P_{\alpha}}$&$P_{\alpha}^*$&${T_{1/2}^\text{pre}}$ (s)\\ \hline
\endfirsthead
\multicolumn{9}{c}%
{{Table 2. -- continued from previous page}} \\
\hline {$\mathcal{\alpha}$ transition}&$Q_{\alpha}$ (MeV)&${j^{\pi}_{p}}\to{j^{\pi}_{d}}$ &$l_{\text{min}}$&$T^\text{exp}_{1/2}$ (s)&${T_{1/2}^\text{cal}}$ (s)&${P_{\alpha}}$&$P_{\alpha}^*$&${T_{1/2}^\text{pre}}$ (s)\\ \hline
\endhead
\hline \multicolumn{9}{r}{{Continued on next page}} \\
\endfoot
\hline
\endlastfoot
$^{251}$Cf $\to^{247}$Cm$^\text{n}$&5.77&${1/2^+}\to{1/2^+}$&0&$2.84\times10^{10}$&$3.38\times10^{10}$&0.42&0.22&$5.50\times10^{10}$\\
$^{253}$Cf $\to^{249}$Cm$^\text{m}$&6.08&${(7/2^+)}\to{(7/2^+)}$&0&$4.96\times10^{8}$&$6.37\times10^{8}$&0.45&0.16&$1.39\times10^{9}$\\
$^{255}$Cf $\to^{251}$Cm $$&5.74&${(7/2^+)}\to{(1/2^+)}$&4&$2.55\times10^{12}$&$2.69\times10^{11}$&0.04&0.12&$7.64\times10^{11}$\\
$^{253}$Es $\to^{249}$Bk $$&6.74&${7/2^+}\to{7/2^+}$&0&$1.77\times10^{6}$&$1.11\times10^{6}$&0.22&0.14&$2.70\times10^{6}$\\
$^{255}$Es $\to^{251}$Bk$^\text{m}$&6.40&${(7/2^+)}\to{7/2^+\#}$&0&$4.28\times10^{7}$&$4.09\times10^{7}$&0.33&0.11&$1.30\times10^{8}$\\
$^{253}$Fm $\to^{249}$Cf$^\text{m}$&7.05&${(1/2)^+}\to{5/2^+}$&2&$2.14\times10^{6}$&$2.24\times10^{5}$&0.04&0.13&$5.97\times10^{5}$\\
$^{257}$Fm $\to^{253}$Cf $$&6.86&${(9/2^+)}\to{(7/2^+)}$&2&$8.68\times10^{6}$&$1.26\times10^{6}$&0.05&0.08&$5.65\times10^{6}$\\
$^{255}$Md $\to^{251}$Es $$&7.91&${(7/2^-)}\to{3/2^-}$&2&$2.28\times10^{4}$&$2.41\times10^{2}$&$3.70\times10^{-3}$&0.09&$9.06\times10^{2}$\\
$^{257}$Md $\to^{253}$Es $$&7.56&${(7/2^-)}\to{7/2^+}$&1&$1.30\times10^{5}$&$3.15\times10^{3}$&0.01&0.07&$1.52\times10^{4}$\\
$^{259}$Md $\to^{255}$Es $$&7.11&${7/2^-\#}\to{(7/2^+)}$&1&$4.43\times10^{5}$&$2.01\times10^{5}$&0.16&0.06&$1.22\times10^{6}$\\
$^{255}$No $\to^{251}$Fm$^\text{m}$&8.23&${(1/2^+)}\to{5/2^+}$&2&$6.92\times10^{2}$&$4.53\times10^{1}$&0.02&0.09&$1.78\times10^{2}$\\
$^{259}$No $\to^{255}$Fm $$&7.85&${(9/2^+)}\to{7/2^+}$&2&$4.62\times10^{3}$&$8.13\times10^{2}$&0.06&0.05&$5.21\times10^{3}$\\
$^{257}$Lr $\to^{253}$Md$^\text{p}$&9.02&${(1/2^-)}\to{1/2^-\#}$&0&$6.00\times10^{0}$&$1.84\times10^{-1}$&0.01&0.07&$9.66\times10^{-1}$\\
$^{259}$Lr $\to^{255}$Md$^\text{p}$&8.58&${1/2^-\#}\to{1/2^-\#}$&0&$7.93\times10^{0}$&$3.95\times10^{0}$&0.17&0.05&$2.62\times10^{1}$\\
$^{257}$Rf$^\text{m}\to^{253}$No $$&9.16&${(11/2^-)}\to{(9/2^-)}$&2&$4.88\times10^{0}$&$2.72\times10^{-1}$&0.02&0.07&$1.44\times10^{0}$\\
$^{259}$Rf $\to^{255}$No$^\text{p}$&9.03&${7/2^+\#}\to{(7/2^+)}$&0&$2.85\times10^{0}$&$3.73\times10^{-1}$&0.05&0.05&$2.50\times10^{0}$\\
$^{261}$Rf $\to^{257}$No $$&8.65&${3/2^+\#}\to{(3/2^+)}$&0&$7.97\times10^{0}$&$5.18\times10^{0}$&0.23&0.04&$4.30\times10^{1}$\\
$^{263}$Rf $\to^{259}$No $$&8.26&${3/2^+\#}\to{(9/2^+)}$&4&$2.20\times10^{3}$&$5.28\times10^{2}$&0.08&0.03&$5.28\times10^{3}$\\
$^{259}$Db $\to^{255}$Lr$^\text{m}$&9.58&${9/2^+\#}\to{(7/2^-)}$&1&$5.10\times10^{-1}$&$2.58\times10^{-2}$&0.02&0.05&$1.71\times10^{-1}$\\
$^{259}$Sg $\to^{255}$Rf $$&9.77&${(11/2^-)}\to{(9/2^-)}$&2&$4.14\times10^{-1}$&$2.51\times10^{-2}$&0.02&0.05&$1.60\times10^{-1}$\\
$^{259}$Sg$^\text{m}\to^{255}$Rf$^\text{m}$&9.71&${(1/2^+)}\to{(5/2^+)}$&2&$2.33\times10^{-1}$&$3.67\times10^{-2}$&0.06&0.05&$2.35\times10^{-1}$\\
$^{261}$Sg $\to^{257}$Rf $$&9.71&${(3/2^+)}\to{(1/2^+)}$&2&$1.86\times10^{-1}$&$3.18\times10^{-2}$&0.06&0.04&$2.54\times10^{-1}$\\
$^{263}$Sg $\to^{259}$Rf $$&9.41&${7/2^+\#}\to{7/2^+\#}$&0&$1.07\times10^{0}$&$1.32\times10^{-1}$&0.04&0.04&$1.28\times10^{0}$\\
$^{265}$Sg $\to^{261}$Rf$^\text{m}$&8.98&${9/2^+\#}\to{9/2^+\#}$&0&$1.84\times10^{1}$&$2.37\times10^{0}$&0.05&0.03&$2.74\times10^{1}$\\
$^{265}$Sg$^\text{m}\to^{261}$Rf $$&9.12&${3/2^+\#}\to{3/2^+\#}$&0&$2.52\times10^{1}$&$8.75\times10^{-1}$&0.01&0.03&$1.01\times10^{1}$\\
$^{271}$Sg $\to^{267}$Rf $$&8.90&${{}{}{}}$&0&$2.66\times10^{2}$&$3.20\times10^{0}$&$4.22\times10^{-3}$&0.02&$5.27\times10^{1}$\\
$^{261}$Bh $\to^{257}$Db$^\text{m}$&10.36&${(5/2^-)}\to{(1/2^-)}$&2&$1.34\times10^{-2}$&$1.45\times10^{-3}$&0.04&0.05&$1.10\times10^{-2}$\\
$^{271}$Bh $\to^{267}$Db $$&9.43&${{}{}{}}$&0&$6.00\times10^{2}$&$1.88\times10^{-1}$&$1.09\times10^{-4}$&0.02&$2.99\times10^{0}$\\
$^{265}$Hs $\to^{261}$Sg $$&10.47&${3/2^+\#}\to{(3/2^+)}$&0&$1.96\times10^{-3}$&$8.53\times10^{-4}$&0.15&0.03&$8.78\times10^{-3}$\\
$^{265}$Hs$^\text{m}\to^{261}$Sg$^\text{m}$&10.60&${9/2^+\#}\to{(11/2^-)}$&1&$3.60\times10^{-4}$&$4.90\times10^{-4}$&0.48&0.03&$5.05\times10^{-3}$\\
$^{267}$Hs $\to^{263}$Sg $$&10.04&${5/2^+\#}\to{7/2^+\#}$&2&$6.88\times10^{-2}$&$1.67\times10^{-2}$&0.09&0.03&$2.01\times10^{-1}$\\
$^{269}$Hs $\to^{265}$Sg $$&9.35&${9/2^+\#}\to{9/2^+\#}$&0&$1.60\times10^{1}$&$8.45\times10^{-1}$&0.02&0.03&$1.15\times10^{1}$\\
$^{273}$Hs $\to^{269}$Sg $$&9.71&${3/2^+\#}\to{{\qquad\quad}}$&0&$1.06\times10^{0}$&$6.32\times10^{-2}$&0.02&0.02&$1.03\times10^{0}$\\
$^{275}$Hs $\to^{271}$Sg $$&9.44&${{}{}{}}$&0&$2.90\times10^{-1}$&$3.45\times10^{-1}$&0.42&0.02&$5.90\times10^{0}$\\
$^{275}$Mt $\to^{271}$Bh $$&10.49&${{}{}{}}$&0&$1.17\times10^{-1}$&$1.10\times10^{-3}$&$3.30\times10^{-3}$&0.02&$1.76\times10^{-2}$\\
$^{267}$Ds $\to^{263}$Hs $$&11.78&${3/2^+\#}\to{3/2^+\#}$&0&$1.00\times10^{-5}$&$3.73\times10^{-6}$&0.13&0.04&$3.69\times10^{-5}$\\
$^{269}$Ds $\to^{265}$Hs$^\text{m}$&11.28&${9/2^+\#}\to{9/2^+\#}$&0&$2.30\times10^{-4}$&$4.08\times10^{-5}$&0.06&0.03&$4.62\times10^{-4}$\\
$^{271}$Ds $\to^{267}$Hs $$&10.88&${13/2^-\#}\to{5/2/^+\#}$&5&$9.00\times10^{-2}$&$3.25\times10^{-3}$&0.01&0.03&$4.10\times10^{-2}$\\
$^{271}$Ds$^\text{m}\to^{267}$Hs $$&10.95&${9/2^+\#}\to{5/2^+\#}$&2&$1.70\times10^{-3}$&$3.46\times10^{-4}$&0.07&0.03&$4.38\times10^{-3}$\\
$^{273}$Ds $\to^{269}$Hs $$&11.38&${13/2^-\#}\to{9/2^+\#}$&3&$2.40\times10^{-4}$&$5.10\times10^{-5}$&0.07&0.03&$7.01\times10^{-4}$\\
$^{273}$Ds$^\text{m}\to^{269}$Hs $$&11.58&${3/2^+\#}\to{9/2^+\#}$&4&$1.20\times10^{-1}$&$3.44\times10^{-5}$&$1.00\times10^{-4}$&0.03&$4.72\times10^{-4}$\\
$^{277}$Ds $\to^{273}$Hs $$&10.83&${11/2^+\#}\to{3/2^+\#}$&4&$6.00\times10^{-3}$&$1.50\times10^{-3}$&0.09&0.02&$2.25\times10^{-2}$\\
$^{279}$Ds $\to^{275}$Hs $$&10.09&${{}{}{}}$&0&$2.10\times10^{0}$&$2.26\times10^{-2}$&$3.77\times10^{-3}$&0.02&$3.42\times10^{-1}$\\
$^{281}$Ds $\to^{277}$Hs $$&9.52&${3/2^+\#}\to{3/2^+\#}$&0&$9.26\times10^{1}$&$8.49\times10^{-1}$&$3.21\times10^{-3}$&0.02&$1.26\times10^{1}$\\
$^{281}$Ds$^\text{m}\to^{277}$Hs$^\text{m}$&9.46&${{}{}{}}$&0&$9.00\times10^{-1}$&$1.28\times10^{0}$&0.50&0.02&$1.90\times10^{1}$\\
$^{279}$Rg $\to^{275}$Mt $$&10.53&${{}{}{}}$&0&$1.80\times10^{-1}$&$3.60\times10^{-3}$&0.01&0.03&$4.90\times10^{-2}$\\
$^{277}$Cn $\to^{273}$Ds$^\text{m}$&11.42&${3/2^+\#}\to{3/2^+\#}$&0&$8.50\times10^{-4}$&$6.13\times10^{-5}$&0.03&0.03&$7.24\times10^{-4}$\\
$^{281}$Cn $\to^{277}$Ds $$&10.46&${3/2^+\#}\to{11/2^+\#}$&4&$1.80\times10^{-1}$&$5.13\times10^{-2}$&0.10&0.03&$6.07\times10^{-1}$\\
$^{285}$Cn $\to^{281}$Ds $$&9.32&${5/2^+\#}\to{3/2^+\#}$&2&$3.20\times10^{1}$&$2.62\times10^{1}$&0.29&0.03&$2.81\times10^{2}$\\
$^{285}$Cn$^\text{m}\to^{281}$Ds$^\text{m}$&9.85&${{}{}{}}$&0&$1.50\times10^{1}$&$4.22\times10^{-1}$&0.01&0.03&$4.52\times10^{0}$\\
$^{283}$Ed $\to^{279}$Rg $$&10.51&${{}{}{}}$&0&$1.60\times10^{-1}$&$1.64\times10^{-2}$&0.04&0.04&$1.63\times10^{-1}$\\
$^{285}$Ed $\to^{281}$Rg $$&10.01&${{}{}{}}$&0&$3.30\times10^{0}$&$3.42\times10^{-1}$&0.04&0.04&$3.19\times10^{0}$\\
$^{285}$Fl $\to^{281}$Cn $$&10.56&${{\qquad\quad}}\to{3/2^+\#}$&0&$2.10\times10^{-1}$&$2.48\times10^{-2}$&0.04&0.04&$1.97\times10^{-1}$\\
$^{287}$Fl $\to^{283}$Cn $$&10.16&${{}{}{}}$&0&$5.20\times10^{-1}$&$2.71\times10^{-1}$&0.18&0.05&$1.99\times10^{0}$\\
$^{289}$Fl $\to^{285}$Cn $$&9.97&${5/2^+\#}\to{5/2^+\#}$&0&$2.40\times10^{0}$&$8.52\times10^{-1}$&0.12&0.05&$5.62\times10^{0}$\\
$^{289}$Fl$^\text{m}\to^{285}$Cn$^\text{m}$&10.17&${{}{}{}}$&0&$1.10\times10^{0}$&$2.33\times10^{-1}$&0.07&0.05&$1.53\times10^{0}$\\
$^{287}$Ef $\to^{283}$Ed $$&10.77&${{}{}{}}$&0&$9.50\times10^{-2}$&$1.43\times10^{-2}$&0.05&0.06&$8.76\times10^{-2}$\\
$^{289}$Ef $\to^{285}$Ed $$&10.52&${{}{}{}}$&0&$3.10\times10^{-1}$&$5.86\times10^{-2}$&0.07&0.06&$3.24\times10^{-1}$\\
$^{291}$Lv $\to^{287}$Fl $$&10.90&${{}{}{}}$&0&$2.80\times10^{-2}$&$1.23\times10^{-2}$&0.15&0.09&$4.92\times10^{-2}$\\
$^{293}$Lv $\to^{289}$Fl $$&10.69&${{\qquad\quad}}\to{5/2^+\#}$&0&$8.00\times10^{-2}$&$3.88\times10^{-2}$&0.17&0.10&$1.34\times10^{-1}$\\
$^{293}$Lv$^\text{m}\to^{289}$Fl$^\text{m}$&10.66&${{}{}{}}$&0&$8.00\times10^{-2}$&$4.67\times10^{-2}$&0.20&0.10&$1.61\times10^{-1}$\\
$^{293}$Eh $\to^{289}$Ef $$&11.30&${{}{}{}}$&0&$2.10\times10^{-2}$&$2.48\times10^{-3}$&0.04&0.13&$6.93\times10^{-3}$\\
$^{293}$Og$\to^{289}$Lv $$&11.92&${1/2^+\#}\to{5/2^+\#}$&2&$1.00\times10^{-3}$&$2.96\times10^{-4}$&0.10&0.16&$6.53\times10^{-4}$\\
$^{295}$Og$\to^{291}$Lv $$&11.70&${{}{}{}}$&0&$1.00\times10^{-2}$&$5.49\times10^{-4}$&0.02&0.19&$1.03\times10^{-3}$\\
$^{297}$Og$\to^{293}$Lv$^\text{m}$&12.00&${{}{}{}}$&0&${}{}{}$&$1.04\times10^{-4}$&{}&0.23&$1.60\times10^{-4}$\\

\end{longtable}
\end{center}

\begin{multicols}{2}
\begin{center}
\includegraphics[width=8.5cm]{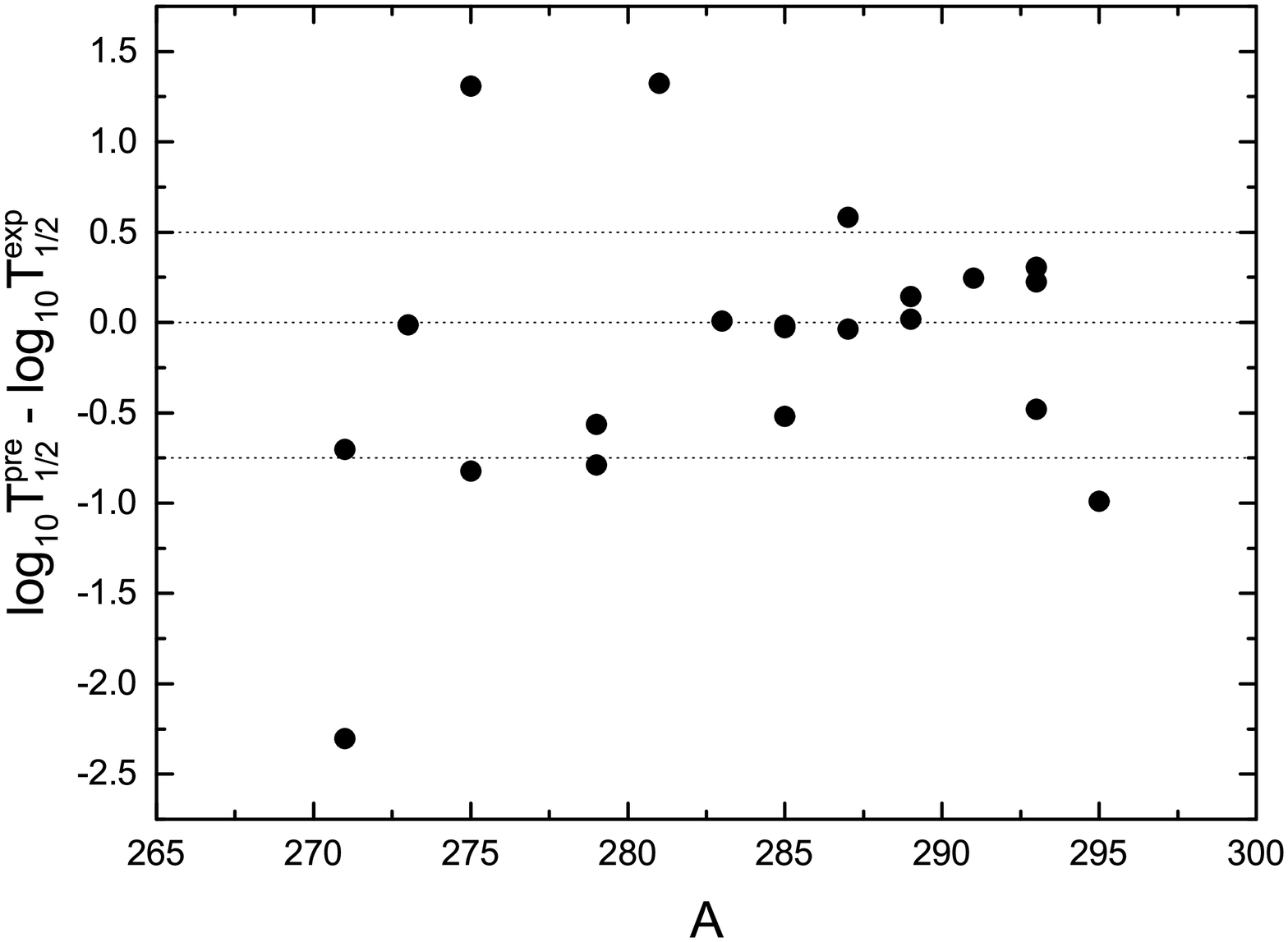}
\figcaption{\label{fig1}The logarithmic differences between $T_{1/2}^\text{pre}$ and $T_{1/2}^\text{exp}$ for $\mathcal{\alpha}$ decay which spin and parity of parent and/or daughter nuclei are unknown and assume $l_{\text{min}}=0$. }
\end{center}

\begin{center}
\includegraphics[width=8.5cm]{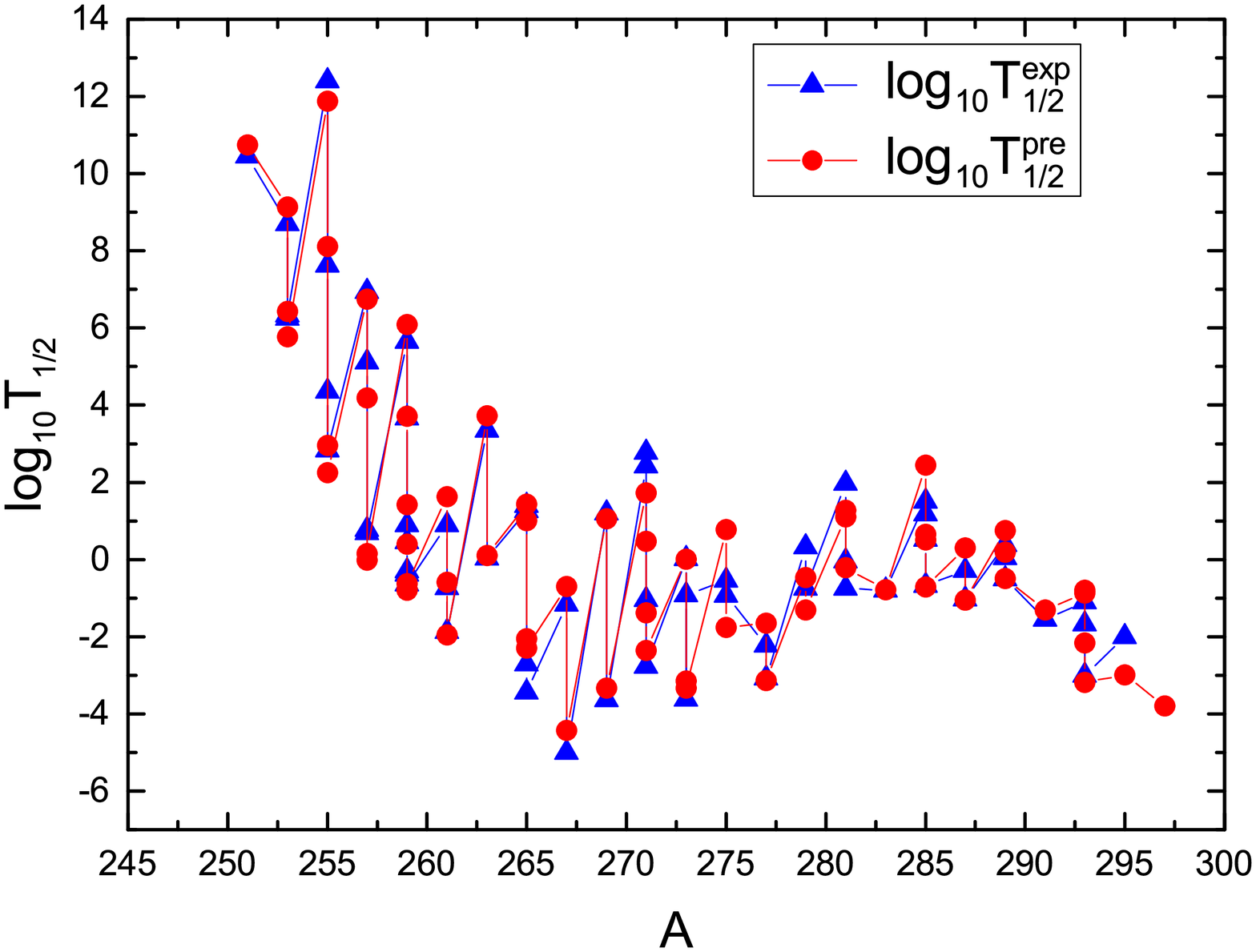}
\figcaption{\label{fig2}(color online) Logarithmic half-lives of experimental data and predicted ones. The blue triangle and red circle denote the experimental half-lives $T_{1/2}^\text{exp}$, and predicted results $T_{1/2}^\text{pre}$, respectively. }
\end{center}

\begin{center}
\includegraphics[width=8.5cm]{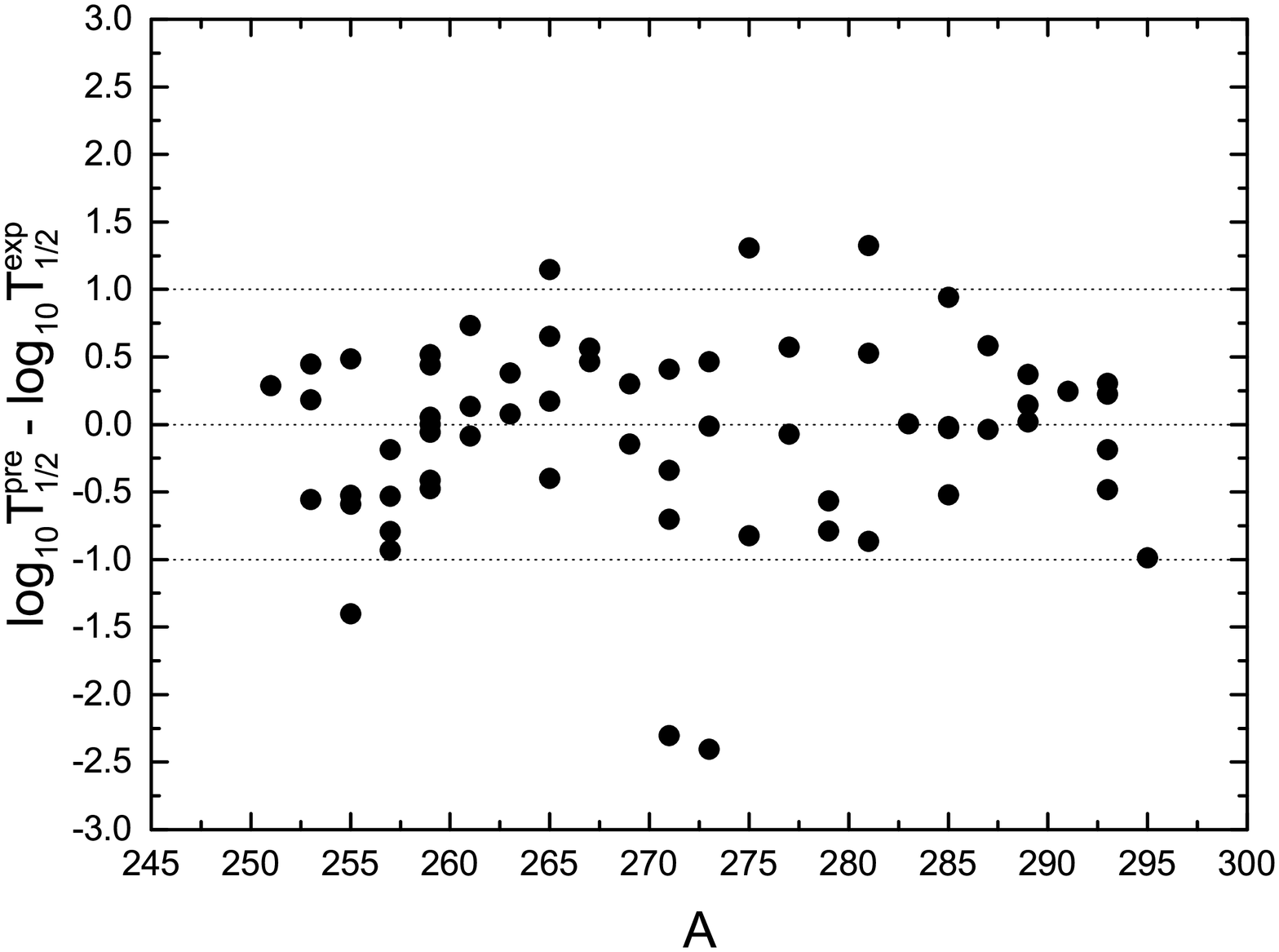}
\figcaption{\label{fig3}The logarithmic differences between $T_{1/2}^\text{pre}$ and $T_{1/2}^\text{exp}$. }
\end{center}

\section{Summary}
In summary, we predict $\mathcal{\alpha}$ decay half-life of $^{297}$Og, and systematically calculate the half-lives of 64 odd-$A$ nuclei in the region of $82<Z<126$ and $152<N<184$ from $^{251}$Cf to $^{295}$Og within TPA, as well as extract corresponding $\mathcal{\alpha}$ preformation probabilities and a new set of parameters for $\mathcal{\alpha}$ preformation probabilities considering the shell effect and proton-neutron interaction. The spin and parity of ground states for nuclei $^{269}$Sg, $^{285}$Fl, $^{293}$Lv are predicted, the corresponding spin and parity may be $3/2^+$, $3/2^+$ and $5/2^+$, respectively.
The predicated $T_{1/2}$ of $^{297}$Og is 0.16 ms within a factor of 4.97. This work will be used as a reference for synthesizing nucleus $^{297}$Og. 
\end{multicols}
\vspace{-1mm}
\centerline{\rule{80mm}{0.1pt}}
\vspace{2mm}

\begin{multicols}{2}

\end{multicols}

\clearpage
\end{CJK*}
\end{document}